\documentclass[11pt]{article}
\usepackage{latexsym}  
\usepackage{amssymb}
\usepackage{graphicx}
\usepackage{amsmath}

\begin{document}

\begin{center}
{\Large\bf Phenomenology of Harmless Family Gauge Bosons\\[.1in]
 to $K^0$-$\bar{K}^0$ Mixing}

\vspace{4mm}
{\bf Yoshio Koide}

{\it Department of Physics, Osaka University, 
Toyonaka, Osaka 560-0043, Japan} \\
{\it E-mail address: koide@kuno-g.phys.sci.osaka-u.ac.jp}

\end{center}

\vspace{3mm}

\begin{abstract}
When we try to consider family gauge bosons with a lower
energy scale, a major obstacle is constraints from the 
observed $P^0$-$\bar{P}^0$ mixings ($P^0=K^0, D^0, B^0,B^0_s$). 
Against such a conventional view, we point out that, in a U(3) 
family gauge boson model, the bosons are harmless 
to any $P^0$-$\bar{P}^0$ mixings independently of explicit values 
of the family mixings, if masses $M_{ij}$ of the gauge bosons 
$A_i^{\ j}$ ($i,j$ are family indexes) satisfy a relation 
$2/M_{ij}^2=1/M_{ii}^2 +1/M_{jj}^2$.  If such the case can be
realized together with an inverted mass hierarchy 
$M_{33}^2 \ll M_{22}^2 \ll M_{11}^2$, we can consider family 
gauge bosons with a considerably lower scale, so that we can 
expect rich signs for family gauge bosons in a TeV scale. 
\end{abstract}


PCAC numbers:  
  11.30.Hv, 
  12.60.-i, 
  14.70.Pw, 

\vspace{5mm}

\noindent{\large\bf 1 \ Introduction} 
 
It seems to be very attractive to understand ``families" 
(``generations") in quarks and leptons from concept of 
a symmetry \cite{f_symmetry}. 
(For a recent work, for instance, see Ref.\cite{Buras}.)
It is also attractive that such family gauge bosons are visible
at our terrestrial energy scale.  
However, when we try to consider such a visible family gauge 
boson model, we always meet with constraints from the observed 
pseudo-scalar-anti-pseudo-scalar ($P^0$-$\bar{P}^0$) 
meson mixings ($K^0$-$\bar{K}^0$, $D^0$-$\bar{D}^0$, and so on).
The constraints are too tight to allow family gauge bosons with 
lower masses, so that it is usually taken that a scale of 
the symmetry braking is considerably high (e.g. an order of 
at least $10^4$ TeV). 
It is usually taken that it is hard to observe gauge boson effects 
even in the LHC era.  
However, if the family gauge symmetry really exists, 
it is rather likely that the effects are certainly visible. 
If we can built a family gauge boson model in which the gauge bosons
do not contribute to the $P^0$-$\bar{P}^0$ mixings, family gauge 
boson effects can become visible at a TeV scale.

Recently, a family gauge boson model \cite{K-Y_PLB12} which 
considerably loosen such the severe constraints from 
the $P^0$-$\bar{P}^0$ mixings have been proposed. 
The model has the following characteristics: 

\noindent
(i) A family gauge symmetry is U(3), so that a number of 
the family gauge bosons are nine (not eight). 

\noindent
(ii) The symmetry breaking is not caused by scalars ${\bf 3}$ and/or 
 ${\bf 6}$ of U(3), but it is caused by a scalar $({\bf 3}, {\bf 3}^*)$ of
U(3)$\times$U(3)$'$, which are broken at $\Lambda$ and
$\Lambda'$ ($\Lambda \ll\Lambda'$), respectively.
Therefore,  a direct gauge boson mixing
$A_i^{\ j} \leftrightarrow A_j^{\ i}$ ($i=1,2,3$) 
does not appear in this model. 

\noindent
(iii) The family gauge boson mass matrix is diagonal 
in a diagonal basis of the charged lepton mass matrix $M_e$.  
Therefore, in the charged lepton sector, the family number is 
exactly conserved.
(Of course, neutrino states which we can observe through 
weak interactions are not $(\nu_1, \nu_2, \nu_3)$, but 
$(\nu_e, \nu_\mu. \nu_\tau)$ which are partners of 
 $(e_L, \mu_L, \tau_L)$. )

\noindent
(iv) In the quark sector, since quark mass matrices $M_u$ and $M_d$ 
are, in general, not always diagonal on the diagonal basis of $M_e$, 
so that family number violations at tree level are caused   
only through the mixing matrices among up- and down-quarks, 
$U^u \neq {\bf 1}$ and  $U^d \neq {\bf 1}$, where eigenstates of 
the family symmetry $(u^0_i, d^0_i)$ are given by 
$(u^0_i, d^0_i)= (U^u_{ij} u_j, U^d_{ij} d_j)$: 
 $$
{\cal H}_{fam} = \frac{g_{F}}{\sqrt{2}} \left[ (\bar{e}_i \gamma_\mu e_j) 
+ (\bar{\nu}_i \gamma_\mu \nu_j) 
+ U^{* u}_{ik} U_{jl}^u (\bar{u}_k \gamma_\mu u_l)
+  U^{* d}_{ik} U_{jl}^d(\bar{d}_k \gamma_\mu d_l) 
 \right] (A_i^{\ j})^\mu .
\eqno(1)
$$
The form is essential to our discussion, so that we give a brief review
of the form (1)
in Appendix. 

\noindent
(v) The gauge boson masses $M_{ij}$ are dominantly generated by vacuum 
expectation values (VEVs) of scalars $\Psi_i^\alpha$ which are 
$({\bf 3}, {\bf 3}^*)$ of U(3)$\times$U(3)$'$, and whose VEVs are given by 
$\langle \Psi_i^\alpha \rangle = \delta_i^\alpha v_i$ as follows:
$$
M^2(A_i^{\ j}) = \frac{1}{2} g_A^2 (|v_i|^2 + |v_j|^2) + \cdots,
\eqno(2) 
$$
where ``$+\cdots$ denotes contributions from other scalars which
are negligibly small, 
so that the family gauge boson masses $M_{ij} \equiv M(A_i^{\ j})$ 
satisfy relations
$$
2 M_{ij}^2 = M_{ii}^2 + M_{jj}^2 .
\eqno(3)
$$
In order to realize the Sumino's cancellation mechanism \cite{Sumino_PLB09}, 
as we discuss later,  
we take an inverted mass hierarchy, $v_i \propto 1/\sqrt{m_{ei}}$, i.e.
$$
M_{ij}^2 \equiv M^2(A_i^j) = k\left( \frac{1}{m_{ei}} +\frac{1}{m_{ej}} \right).
\eqno(4)
$$
Therefore, Eq.(4) gives family gauge boson masses with an inverted 
mass hierarchy $M_{33} \ll M_{22} \ll M_{11}$.
Here, note that the scalar $\Psi$ is different from a scalar $\Phi$ 
which generates charged lepton masses $m_{ei}$.
Since the model gives a VEV relation 
$\langle \Psi \rangle \langle \Phi^\dagger \rangle \propto {\bf 1}$,
the gauge boson mass matrix is diagonal when the charged lepton 
mass matrix is diagonal. 
Also note that Eq.(4) is an approximate relation under 
$|\langle \Psi\rangle |^2 \gg |\langle \Phi \rangle|^2$.  
(For more details, see Appendix, Eq.(A.5).)

The model with the inverted mass hierarchy (K-Y model \cite{K-Y_PLB12}) is 
an extended version of the Sumino model \cite{Sumino_PLB09}.
In the Sumino model, the gauge coupling constant $g_F$ is not free 
parameter.  
Sumino has paid 
why the charged lepton mass relation \cite{K-mass,K-mass2,K-mass3}
$$ 
m_e + m_\mu +m_\tau = \frac{2}{3} \left(\sqrt{m_e}+ \sqrt{m_\tau} 
+\sqrt{m_\tau}\right)^2, 
\eqno(5)
$$  
is well satisfied by the pole masses (not by the running masses). 
The running masses $m_{ei}(\mu)$ are given by \cite{Arason92}
$$
m_{ei}(\mu) = m_{ei} \left[ 1-\frac{\alpha_{em}(\mu)}{\pi} 
\left( 1 +\frac{3}{4} \log \frac{\mu^2}{m_{ei}^2(\mu)} \right) \right] .
\eqno(6)
$$
If the factor $\log(m_{ei}^2/\mu^2)$ in Eq.(6) is absent, then the 
running masses $m_{ei}(\mu)$ are also satisfy the formula (5).
Sumino has required that contribution of family gauge bosons to the 
charged lepton mass $m_{ei}(\mu)$ cancels the factor $\log(m_{ei}^2/\mu^2)$
due to photon.   
(However, in the present paper, we do not require Sumino's cancellation 
mechanism, so that we do not refer the details.) 
In the K-Y model, too, the coupling constant $g_F$ is not a free parameter 
in the model.
In order to cancel a factor $\log (m_{ei}^2/\mu^2)$ in the QED correction
by a factor $\log (M_{ij}^2/\mu^2)$ due to family gauge boson exchanges, 
the gauge boson masses must have inverted masses 
$M^2_{ii} \propto m_{ei}^{-1}$.
Therefore, the characteristic (v) in the K-Y model, ``family gauge bosons 
with an inverted mass hierarchy", is not an assumption, 
but inevitable consequence of the model.  
However, in this paper, we do not require the cancellation mechanism, so that 
the characteristic (v) is a phenomenological assumption 
in the present scenario.

However, even in the K-Y model, it is still difficult to reduced the lightest 
gauge boson mass to a few TeV energy scale \cite{YK_PRD13}.  
In Sec.2, we point out that if masses $M_{ij}$ of the family gauge bosons 
$A_i^{\ j}$ satisfy a relation
$$
\frac{2}{M_{ij}^2} =\frac{1}{M_{ii}^2} +\frac{1}{M_{jj}^2} ,
\eqno(7)
$$
the family gauge bosons do not contribute to the  $P$-$\bar{P}$ 
mixings at all.
Of course, such the mechanism based on the relation (7) is
effective only in a model in which there is no direct transition
$A_i^{\ j} \leftrightarrow A_j^{\ i}$, i.e. in which gauge bosons interact 
with quarks and leptons according to Eq.(1).

The purpose of the present paper is to discuss visible effects of the family 
gauge bosons at TeV scale under the assumption (7) from the phenomenological
point of view, but not to build a model with the mass relation (7) 
from the theoretical point of view. 
In Sec.2, we demonstrate that the family gauge boson cannot contribute to 
the $P^0$-$\bar{P}^0$ mixing at all when we assume the mass relation (7). 
In Sec.3, phenomenological investigation is given under the assumption (7). 
We speculate that $M_{33}/(g_F/\sqrt{2}) \sim 5.1$ TeV and 
$M_{23}/(g_F/\sqrt{2}) \simeq M_{31}/(g_F/\sqrt{2})\sim 7.3$ TeV,
while $M_{12}/(g_F/\sqrt{2}) \sim 500$ TeV. 
(In the present model, differently from the Sumino model and the K-Y model, 
we cannot fix the exact values of $M_{ij}$, 
since $g_F$ is free parameter.)
If $g_F$ is $g_F/\sqrt{2}\sim 0.2$, we can guess that 
$M_{33}$, $M_{23}$ and $M_{31}$ are of an order of $1$ - $2$ TeV, 
so that we are able to observe those at the LHC with $\sqrt{s}=14$ TeV via 
$A_3^{\ 3} \rightarrow \tau^+ \tau^-$, $A_3^{\ 2} \rightarrow \mu^+ \tau^-$
and  $A_3^{\ 1} \rightarrow e^+ \tau^-$.  
The value $M_{12}/(g_F/\sqrt{2}) \sim 500$ TeV is within our reach of 
our observation of $\mu$-$e$ conversion in the near future experiments. 
 Especially, an observation of $\mu^- N \rightarrow e^- N$ 
(but non-observation of $\mu \rightarrow e +\gamma$) will 
be a promising as a test of the present scenario. 
Finally, Sec.4 is devoted to concluding remarks.

\vspace{3mm}

\noindent{\large\bf 2 \ Harmless condition to $P$-$\bar{P}$ mixings }

We start from the family gauge boson interactions given in Eq.(1).  
The interactions (1) can be derived, for example, from 
a model U(3)$\times$U(3)$'$ mode (see Appendix). 
Then, we can express effective quark current-current interactions
with a family number change $\Delta N_{fam} =2$ as follows:
$$
H^{eff} = \frac{1}{2} g_F^2 \left[ \sum_i \frac{ (\lambda_i)^2 }{M_{ii}^2} 
+ 2 \sum_{i<j} \frac{\lambda_i \lambda_j }{M_{ij}^2} \right]
(\bar{q}_k \gamma_\mu q_l) (\bar{q}_k \gamma^\mu q_l ) 
\eqno(8)
$$
where
$$
\lambda_1 = U^*_{1k} U_{1l} ,  \ \ \lambda_2 = U^*_{2k} U_{2l} ,  \ \ 
\lambda_3 = U^*_{3k} U_{3l} .  
\eqno(9)
$$
(For example, for a case of $K^0$-$\bar{K}^0$ mixing are given by
 $\lambda_1=U^{d*}_{11} U^d_{12}$, $\lambda_2= U^{d*}_{21} U^d_{22}$ 
and $\lambda_3= U^{d*}_{31} U^d_{32}$.)
These $\lambda_i$ with $k\neq l$ satisfy a unitary triangle condition
$$
\lambda_1 +\lambda_2 + \lambda_3 = 0 .
\eqno(10)
$$
We define the effective coupling constant $G^{eff}$ in the current-current
interaction as 
$$
G^{eff} = \frac{1}{2} g_F^2 \left[ \frac{\lambda_1^2}{M_{11}^2} +
 \frac{\lambda_2^2}{M_{22}^2} +  \frac{\lambda_3^2}{M_{33}^2} +
 2\left( \frac{\lambda_1 \lambda_2 }{M_{12}^2} +
 \frac{\lambda_2 \lambda_3 }{M_{23}^2} + \frac{\lambda_3 \lambda_1 }{M_{31}^2}
 \right) \right] .
\eqno(11)
$$
Obviously, a case of $M_{11}=M_{22}=M_{33}=M_{12}=M_{23}=M_{31}$ gives 
$G^{eff}=0$, because of 
$G^{eff} \propto (\lambda_1 + \lambda_2 + \lambda_3)^2$. 
However, the case is not attractive phenomenologically.

Another case which can give $G^{eff}=0$ is a case with 
the relation (7).
In fact, the effective coupling constant $G^{eff}$ under 
the relation (7) is expressed as
$$
G^{eff} = \frac{1}{2} g_F^2 \left[ \sum_i \frac{ (\lambda_i)^2 }{M_{ii}^2} 
+ \sum_{i<j} \lambda_i \lambda_j\left( \frac{1}{M_{ii}^2} +
 \frac{1}{M_{jj}^2} \right) \right] 
= \frac{1}{2} g_F^2 (\lambda_1 +\lambda_2 +\lambda_3)
\left(\frac{ \lambda_1 }{M_{11}^2} +\frac{ \lambda_2 }{M_{22}^2} 
+ \frac{ \lambda_3 }{M_{33}^2} \right) ,
\eqno(12)
$$ 
so that, because of the unitary triangle condition (10), we can obtain  
$G^{eff}=0$ for any values of the quark mixing.

However, note that if we consider that the U(3) family symmetry is broken
by a scalar ${\bf 6}$ (and/or ${\bf 6}^*$), we cannot prevent the 
$P$-$\bar{P}$ mixing even with the mass relation (7), because, in such a 
case, $A_i^{\ j}$-$A_j^{\ i}$ mixing directly appears via vacuum 
expectation value (VEV) of the scalar ${\bf 6}$ (and/or ${\bf 6}^*$). 
In the K-Y model, the U(3) symmetry is broken only by the scalar 
$({\bf 3},{\bf 3}^*)$ of U(3)$\times$U(3)$'$, so that 
the effective interactions with $\Delta N_{fam}=2$ are caused only 
by Eq.(3). 
(This suppression mechanism is a kind of the GIM mechanism \cite{GIM_PRD70}.
This is peculiar to the quark current-current interactions with 
$\Delta N_{fam}=2$, and it does not work in quark-lepton interaction
with $\Delta N_{fam}=1$.)

In general, since we have six gauge boson masses and three constraints (7), 
we can describe five gauge boson mass ratios by two parameters.
We define parameters $a$ and $b$ as
$$
a \equiv \frac{M_{22}}{M_{33}} , \ \ \ \ 
b \equiv \frac{M_{11}}{M_{33}} . 
\eqno(13)
$$
If we assume an inverted mass hierarchy with $b^2 >a^2 > 1$, 
we obtain the gauge boson mass ratios as follows:
$$
M_{33} : M_{23} : M_{31} : M_{22} : M_{12} : M_{11} =
1 : \sqrt{ \frac{2}{1+1/a^2}} :  \sqrt{ \frac{2}{1+1/b^2}} :  
a : \sqrt{ \frac{2}{1+a^2/b^2}}\, a :  b ,
\eqno(14)
$$
which leads to
$$
M_{33} : M_{23} : M_{31} : M_{22} : M_{12} : M_{11} \simeq 
1 : \sqrt{2} :  \sqrt{2}  :  
a : \sqrt{2 } a :  b ,
\eqno(15)
$$
under the assumption $b^2 \gg a^2 \gg 1$. 
If we can give two parameters $a$ and $b$, then we can fix  
all the gauge bosons mass ratios. 
(Note that the parameters $a$ and $b$ are fixed by charged 
lepton mass ratios in the Sumino model and the K-Y model
in which the gauge boson masses satisfy the relation (3),
while the parameters $a$ and $b$ in the present model 
are free parameters. )

\vspace{3mm}

\noindent{\large\bf 3 \ Phenomenology of the family gauge bosons}

In this section, we discuss phenomenology of the family gauge bosons
whose masses satisfy the harmless condition (7).
Let us forget about the theoretical origin of mass spectrum (7) 
for the time being.
We optimistically consider that the relation will be derived by 
considering  a scalar $({\bf 6},{\bf 6}^*)$ of U(3)$\times$U(3)$'$ 
and/or a mixing with another gauge bosons (for example,  
[U(1)]$^3$ gauge bosons).
In order to investigate the origin of the relation (7) in the near future, 
it is important to investigate phenomenological aspect. 

{\bf 3.1 \  Constraints from rare $K$ and $B$ decay searches}

First, let us see experimental lower limit of the family gauge bosons. 
We do not need an explicit value of $g_F$
as far as we discuss phenomenon due to the current-current interactions. 
It is convenient to define
$$
\tilde{M}_{ij}^2 \equiv  \frac{M_{ij}^2}{g_F^2/2} .
\eqno(16)
$$
As far as we treat four-Fermi current-current interactions,
the value $\tilde{M}_{ij}$ are practically useful rather 
than $M_{ij}$. 
Real mass values $M_{ij}$ are needed only when we discuss a direct 
observation of $A_i^j$ (for example, $pp \rightarrow
A_3^{\ 3} +X \rightarrow \tau^+\tau^- X$).  
 
For example, we can estimate a rare decay 
$K^+ \rightarrow \pi^+ e^- \mu^+$ as follows:
$$
\frac{Br(K^+ \rightarrow \pi^+ e^- \mu^+)}{
Br(K^+ \rightarrow \pi^0 \mu^+ \nu_\mu)} = (r_{12})^4 
\frac{|U_{11}^{* d} U_{22}^d |^2}{|V_{us}|^2}
\frac{ f(m_{\pi^+}/m_K)}{\frac{1}{2}f(m_{\pi^0}/m_K)} \simeq 
\frac{2}{|V_{us}|^2} (r_{12})^4 ,
\eqno(17)
$$
where
$$
(r_{ij})^2 =  \frac{ (g_F^2/2)/M_{ij}^2 }{(g_w^2/8)/M_W^2} 
= \frac{2 v_H^2}{\tilde{M}^2_{12}} ,
\eqno(18)
$$
$v_H = 246\ {\rm GeV}$, and $f(x)$ is a phase space function 
$f(x)= 1-8 x^2 + 8 x^6 -x^8 -12 x^4 \log x^2$. 
(We have neglected the lepton masses.)
Note that the weak interactions are $V-A$, while our family 
gauge boson interactions are pure $V$.
The present data \cite{PDG12} show 
$Br(K^+ \rightarrow \pi^0 \mu^+ \nu_\mu) =(3.353\pm 0.034)$ \% 
and $Br(K^+ \rightarrow \pi^+ e^- \mu^+) <1.3 \times 10^{-11}$,
so that we obtain a lower limit of $\tilde{M}_{12} \sim 196$ TeV.
We show results of lower limits of $\tilde{M}_{ij}$
from the observed rare pseudo-scalar meson decays in Table 1.

\begin{table}[h]
\caption{
Experimental lower limit of $\tilde{M}_{ij} \equiv M_{ij}/(g_F/\sqrt{2})$.
(For $K^+ \rightarrow \pi^+ \nu \bar{\nu}$, see the text.)}
\begin{tabular}{|cc|cc|}\hline
 & Input &  & Output [TeV] \\ \hline
 $Br(K^+ \rightarrow \pi^+ e^- \mu^+)$ & $< 1.3 \times 10^{-11}$ &
$\tilde{M}_{12}$ & $> 196$ \\ 
  $Br(K_L \rightarrow \pi^0 e^\mp \mu^\pm)$ & $< 7.6 \times 10^{-11}$ &
$\tilde{M}_{12}$ & $> 151 $ \\
  $Br(K_L \rightarrow \pi^0 \nu \bar{\nu})$ & $< 2.6 \times 10^{-8}$ &
$\tilde{M}_{12}$ & $> 17.5 $ \\
  $Br(K^+ \rightarrow \pi^+ \nu \bar{\nu})$ & $ (1.7\pm 1.1)\times 10^{-10}$ &
$\tilde{M}_{12}$ & $ \sim 243 $ \\ \hline
 $Br(B^+ \rightarrow K^+ \mu^\pm \tau^\mp)$ & $< 7.7 \times 10^{-5}$ &
$\tilde{M}_{23}$ & $> 4.11$ \\
 $Br(B^+ \rightarrow K^+ \nu \bar{\nu}))$ & $< 1.3 \times 10^{-5}$ &
$\tilde{M}_{23}$ & $> 5.4$ \\
 $Br(B^0 \rightarrow K^0 \nu \bar{\nu})$ & $< 5.6 \times 10^{-6}$ &
$\tilde{M}_{23}$ & $> 6.7 $ \\
 \hline
 $Br(B^0 \rightarrow \pi^0 \nu \bar{\nu})$ & $< 2.2 \times 10^{-4}$ &
$\tilde{M}_{31}$ & $> 4.8 $ \\ \hline
\end{tabular}
\end{table}
 
In Table 1, only $Br(K^+ \rightarrow \pi^+ \nu \bar{\nu})$
has been reported with a finite value of the branching ratio,
$(1.7 \pm 1.1) \times 10^{-10}$ \cite{PDG12}. 
It is usually taken that this value is consistent with
the standard model prediction 
\cite{B-Kpinunu,B-Kpinunu2}
$$
Br(K^+ \rightarrow \pi^+ \nu \bar{\nu})_{SM} = (0.80 \pm 0.11)
\times 10^{-10} .
\eqno(19)
$$
Since our purpose is to find a room for new physics as much as possible,
we take the center value of the observed value.
Then, we can obtain a value $\tilde{M}_{12} \sim 243$ TeV 
shown in Table 1. 
Therefore, exactly speaking, the value $\tilde{M}_{12} \sim 243$ TeV 
should be regarded as a lower limit of the mass of the family
gauge boson $A_2^{\ 1}$.  
(Here, we have regard 
$Br(K^+ \rightarrow \pi^+ \nu \bar{\nu}) \simeq 
Br(K^+ \rightarrow \pi^+ \nu_e \bar{\nu}_\mu)$. )
Also note that our gauge bosons interact with fermions
as a pure vector, while  they behave as 
V-A for decay into neutrinos, because $\nu_R$ are extremely heavy.

As seen in Table 1, the data roughly show 
$\tilde{M}_{12} \geq 250$ TeV and $\tilde{M}_{23} \geq 7$ TeV. 
If we want a model in which contains a family gauge boson with 
a TeV scale mass, it seems to be better to consider a family 
gauge boson model with an inverted mass hierarchy.   

Since we consider that the family gauge boson mass matrix is diagonal
in the diagonal bases of the charged lepton mass matrix, it is likely
that the gauge boson masse ratios are described by the charged lepton 
mass ratios as in the Sumino model and the K-Y model. 
Suggested by the K-Y model, by a way of trial, let us assume
that the parameters $a$ and $b$ defined by Eq.(13) are given by
$$
a= \left(\frac{1/m_\mu}{1/m_\tau}\right)^{n/2} , \ \ \ \ 
b= \left( \frac{1/m_e}{1/m_\tau}\right)^{n/2} . 
\eqno(20)
$$
In the K-Y model \cite{K-Y_PLB12}, a case of $n=1$ in Eq.(20) was adopted.  
However, it has been demonstrated tat the K-Y model with $n=1$ cannot give 
a family gauge boson with a TeV scale from a phenomenological study 
in Ref.\cite{YK_PRD13}.
In the K-Y model, the case $n=1$ has been derived by considering 
$\langle \Phi \rangle \langle \bar{\Psi} \rangle \propto {\bf 1}$, where 
$\Phi$ and $\Psi$ are scalars $({\bf 3},{\bf 3}^*)$ of 
U(3)$\times$(3)$'$, as given a review in Appendix.
Note that we cannot make a singlet state $({\bf 1}, {\bf 1})$ from
three  $({\bf 3},{\bf 3}^*)$, i.e. $\Phi \bar{\Phi} \Psi$.
A possible case of $({\bf 1}, {\bf 1})$ with a next smaller $n$ 
is $\Phi \bar{\Phi} \Phi \bar{\Psi}$, i.e. a case of $n=3$. 
Therefore, in the present model, we consider the case of $n=3$:  
$a=68.96$ and $b=2.050\times 10^5$. 
Although we have speculated $\tilde{M}_{12} \sim 250$ TeV 
in Table 1 from the observed value of 
$Br(K^+ \rightarrow \pi^+ \nu \bar{\nu})$, we conservatively take 
a value of $\tilde{M}_{12}\sim 500$ TeV in the present estimate.
Then, we can speculate 
$$
\tilde{M}_{33} \sim 5.1\  {\rm TeV}, \ \ \
\tilde{M}_{23} \simeq \tilde{M}_{31} \sim 7.3\ {\rm TeV}, \ \ \
\tilde{M}_{22} \sim 350 \  {\rm TeV}, \ \ \ \tilde{M}_{12} \sim 500\  
{\rm TeV}, \ \ \  \tilde{M}_{11} \sim 1.1\times 10^6 \ {\rm TeV}, 
\eqno(21)
$$
The predicted mass values $\tilde{M}_{ij}$ do not conflict with 
the lower limits of the observed values given in Table 1.
Obviously, the gauge boson $A_1^{\ 1}$ is invisible.
However, $A_3^{\ 3}$ and $A_2^{\ 1}$ have possibility 
to be observed at the LHC and at the COMET experiment \cite{Comet}, 
respectively.

 
{\bf 3.2 \  Other visible family gauge boson effects}

Let us discuss possible visible effects of the family gauge bosons
with the mass spectrum (21). 

\noindent
(i) {\it Deviation from the $e$-$\mu$-$\tau$ universality in tau decays} 

Previously, we have estimated \cite{YK_PRD13} a mass value of $\tilde{M}_{23}$ 
as $\tilde{M}_{23} = 5.2^{+6.4}_{-1.4}$ TeV,
from the deviation $\delta =0.0020\pm 0.0016$ in 
$Br(\tau^- \rightarrow \mu^- \nu \bar{\nu}/e^- \nu \bar{\nu})$. 
(In Ref. \cite{YK_PRD13}, the result has been represented in terms of 
$M_{ij}$, in which $g_F/\sqrt{2}= 0.4999$ has been taken.) 
Regrettably, we cannot extract such the value in the present 
model, because the previous value was extracted under an assumption  
$\tilde{M}_{23}^2 \ll \tilde{M}_{31}^2$, while since the mass 
spectrum in the present model gives 
$\tilde{M}_{23}^2 \simeq \tilde{M}_{31}^2$, the previous value 
$\tilde{M}_{23} \sim 5.2$ TeV cannot be derived from the present model. 

On the other hand, we can see sizable deviations from the 
$e$-$\mu$-$\tau$ universality in the $\Upsilon$ decays,
$\Upsilon \rightarrow \tau^+ \tau^- / \mu^+ \mu^- /e^+ e^-$.
We have estimated \cite{YK_PRD13} a mass value of $M_{33}$
as $\tilde{M}_{33} = 0.22^{+0.26}_{-0.05}$ TeV. 
However, the previous value of $\tilde{M}_{33}$ is too small 
compared with the value given in (21).  
However, note that upper value in the previous estimate contain infinity 
if we take 1.3 $\sigma$ of the observed deviation. 
Therefore, the previous result is not conflict with the present 
estimate in (21).
The value of $\tilde{M}_{33}$ given in (21) will be confirmed 
in the $\Upsilon$ decay in the near future.

\noindent
(ii) {\it Lepton number violating rare decays of $B$ and $K$}

For lepton-flavor violating rare decays of $B$ and $K$, 
$B$ decays, $B^+ \rightarrow K^+ \mu^- \tau^+$ and,  
$B^0 \rightarrow K^+ \mu^- \tau^+$, and $K$ decays, 
$K^+ \rightarrow \pi^+ \mu^- \tau^+$, $K_L \rightarrow \pi^0 \nu \bar{\nu}$
and $K^+ \rightarrow \pi^+ \mu^\mp \tau^\pm$ 
 become soon within our reach as seen in Table 1.

\noindent
(iii) {\it $\mu$-$e$ conversion} 

Most sensitive test for our scenario is to observe the so-called 
$\mu$-$e$ conversion. 
(For a review of the $\mu$-$e$ conversion and  more detailed calculations,
for example, see Ref.\cite{Kuno-Okada_RMP01} and 
Ref.\cite{Kitano_PRD02}, respectively.) 
At present, we do not know values of $|U_{11}^{q*} U_{21}^q|$ ($q=u,d$).
Therefore, it is not practical, at this stage, to estimate a $\mu$-$e$ 
conversion rate strictly. 
Instead, we roughly estimate a $\mu$-$e$ conversion rate
in the quark level as follows: 
$$
R_q \equiv \frac{ \sigma(\mu^- + q \rightarrow e^- + q)}{
\sigma( \mu^- + u \rightarrow \nu_\mu + d) }
\simeq \left( \frac{|U_{11}^{q*} U_{21}^q|}{|V_{ud}|} 
\frac{g_F^2/2}{\tilde{M}_{12}^2} 
\frac{M_W^2}{g_w^2/8} \right)^2 = 
\left( \frac{|U_{11}^{q*} U_{21}^q|}{|V_{ud}|} 
(r_{12})^2 \right)^2,
\eqno(22)
$$
where $q=u, d$, and $(r_{12}^2)$ is defined by Eq.(18). 
It is likely that $|U_{21}^u|^2\ll |U_{21}^d|^2$.  
Then, we may regard 
the ratios $R_q$ as $R_u \ll R_d$, so that we can neglect contribution 
to nucleon from $R_u$ compared with that from $R_d$. 
When we suppose 
${|U_{11}^{d*} U_{21}^d|}/{|V_{ud}|} \sim |V_{cd}| \sim 10^{-1}$,  
we can roughly estimate values of $R_d$ for the input values 
 $\tilde{M}_{12} \sim 500$ TeV as $R_d \sim 0.95 \times 10^{-14}$.
Present experimental limit is, for instance for $Au$,  $R(Au) \equiv
\sigma(\mu^- +Au \rightarrow e^- + Au)/\sigma(\mu\ {\rm capture}) 
< 7\times 10^{-13}$ \cite{SINDRUM06}.
The estimated values $R_d \sim 10^{-14}$ become within 
reach of our observation. 
(Although the estimated value $R_d$ has different physical meaning  
from the value $R(Au)$, we consider that the order of the value $R_d$
can provide one with useful information.)
Since the decay $\mu^- \rightarrow e^- +\gamma$ is highly 
suppressed in the present scenario, if we observe 
$\mu^- N \rightarrow e^- N$ without observation of 
$\mu^- \rightarrow e^- +\gamma$, 
then it will strongly support our family gauge boson scenario.
(The decay $\mu^- \rightarrow e^- +\gamma$ can occur through a 
quark-loop diagram.
However, such a diagram is highly suppressed.) 

\noindent
(iv) {\it Direct production of the light gauge bosons 
$A_3^{\ 3}$, $A_3^{\ 2}$ and $A_3^{\ 1}$}  

It should be noted that the values $\tilde{M}_{33} \simeq 5.1$ TeV 
and $\tilde{M}_{32} \simeq \tilde{M}_{31} \simeq 7.2$ TeV
are not real mass values of these family gauge bosons.
The observed masses $M_{ij}$ are given by 
$M_{ij} = (g_F/\sqrt{2}) \tilde{M}_{ij}$. 
If the gauge coupling constant $g_F$ is $g_F/\sqrt{2} \sim 0.2$,
direct productions at the LHC will become hopeful. 
Then, we can observe typical decay modes 
$A_3^{\ 3} \rightarrow \tau^- \tau^+$, 
$A_3^{\ 2}\rightarrow \tau^- \mu^+$ and 
$A_3^{\ 1}\rightarrow \tau^- e^+$ 
with the branching fraction 2/15=13.3\%.
(For example, in the decays of $A_3^{\ 2}$, we have
decay modes, $t+\bar{c}$, $b+\bar{s}$, $\tau^- + \mu^+$ 
and $\nu_\tau + \bar{\nu}_\mu$ with branching fractions 
6/15, 6/15, 2/15 and 1/15, respectively.)

Meanwhile, note that our family gauge interactions are only 
interactions which can interact not only with $\nu_L$ 
but also with $\nu_R$.
The branching ratio $Br(A_i^{\ j} \rightarrow \nu_i \bar{\nu}_j) 
=1/15=6.7 \%$ is one in the case of Majorana neutrinos.
If neutrinos are Dirac neutrinos, the branching ratios is given 
$Br(A_i^{\ j} \rightarrow \nu_i \bar{\nu}_j) =2/16=12.5 \%$.
Therefore, in future, when the data of the direct production of 
$A_i^{\ j}$ are accumulated, we will be able to conclude 
whether neutrinos are Dirac or Majorana by observing whether 
$Br(A_i^{\ j} \rightarrow \nu_i \bar{\nu}_j)$ is $6.7\%$ or $12.5\%$. 

Finally, we would like to comment on decays of pseudo-scalar mesons 
$K^0$, $D^0$, $B_d^0$ and $B_s^0$ into dilepton pair $\ell^+ \ell^-$.
Since the coupling types in the K-Y model \cite{K-Y_PLB12} is 
pure vector type as seen in Eq.(1), 
the gauge bosons cannot couple to a single gauge boson with $J^P=0^-$.
Therefore, the K-Y family gauge bosons cannot contribute to 
$0^- \rightarrow \ell^+ \ell^-$ (and also $\ell^+ \ell^{\prime -}$) 
decays.  
(Also note that the present family gauge bosons cannot contribute to
$P^0$-$\bar{P}^0$ mixing in the $s$ cannel. This is one of reason that 
the contributions of the  K-Y family gauge bosons to $P^0$-$\bar{P}^0$ 
mixing is smaller than the conventional family gauge boson model.) 
On the other hand, family gauge bosons in the original Sumino model
\cite{Sumino_PLB09} have $(V\pm A)$ interactions, the Sumino gauge 
bosons can be distinguished from the K-Y bosons by these $0^-$ meson decays.
Of course, since three body decay discussed in Eq.(19) are caused by 
vector current-current interactions, we cannot apply to the same 
argument as $0^- \rightarrow \ell^+ \ell^-$ to $0^- \rightarrow 0^- 
\ell \bar{\nu}$ decays.

\vspace{3mm}

\noindent{\large\bf 4 \ Concluding remarks}

We have pointed out that if family gauge boson masses satisfy the relation (7),
the gauge bosons are harmless to $P^0$-$\bar{P}^0$ mixing. 
This is valid only in a model in which 
there is no direct $A_i^{\ j} \leftrightarrow A_i^{\ j}$ transition  
and the family number violation is caused only by quark mixing 
($U^d \neq {\bf 1}$ and/or $U^u \neq {\bf 1}$). 

We would like to emphasize the mass relation (7) is promising
from a phenomenological point of view.
If we had adopted the relation (3) instead of the relation (7),
we would obtain a mass relation
$$
M_{33} : M_{23} : M_{22} : M_{31} : M_{12} : M_{11} =
$$
$$
1 : \sqrt{ \frac{1+a^2}{2}} : a :  \sqrt{ \frac{1+b^2}{2}} :  
 \sqrt{ \frac{a^2+b^2}{2}} :  b 
 \simeq 
1 : \frac{a}{\sqrt{2}} : a: \frac{b}{\sqrt{2}} : \frac{b}{\sqrt{2}} : b ,
\eqno(23)
$$
under the assumption $b^2 \gg a^2 \gg 1$, instead of the relation (14). 
As seen by comparing the relation (15) with (23), we can have three 
light bosons $A_3^{\ 3}$, $A_2^{\ 3}$ and $A_1^{\ 3}$ and two bosons 
 $A_2^{\ 2}$ and $A_1^{\ 2}$ with masses of the order of $a \, M_{33}$ 
in the model with the relation (7), while, in the model with the relation (3), 
we have only one lightest boson $A_3^{\ 3}$ and two bosons $A_2^{\ 3}$ and 
$A_2^{\ 2}$ with masses of the order of $a \, M_{33}$. 
Therefore, the model with relation (3) cannot give any interesting 
phenomenology.

However, note that the mechanism of the family symmetry breaking
which is caused by a scalar $({\bf 3}, {\bf 3}^*)$ of
U(3)$\times$U(3)$'$ can give the gauge boson interactions (1),
but, at the same time, the mechanism leads to the mass 
relation (3). 
Regrettably, we do not know  mechanism which gives the 
interaction (1) and also gives the mass relation (7). 
Our next task is to derive the relation (7). 

Meanwhile, note that 
we do not need to require that the relation (7) should exactly 
be satisfied.
The relation (7) may approximately be satisfied in practice.
For example, when we suppose $M_{22} \simeq M_{12} \simeq M_{11}$,
the family gauge boson contribution to $\Delta m_K$ (also 
$\Delta m_D$) can become negligibly small.
The relation (7) is a very useful measure to model-builders 
who consider a family gauge boson model with a lighter scale.  

Anyhow, if the relation (7) is satisfied, at least, approximately,
we can speculate many fruitful family gauge boson effects under 
the mass relation (7).
However, the relation (7) is purely phenomenological one. 
Especially, the numerical values in Eq.(21) should not be taken rigidly.
The values are highly dependent on the tentative input 
$\tilde{M}_{12} \simeq 500$ TeV.
The purpose of the present paper is to point out
a possibility that masses of the family gauge bosons 
are considerably small, and it is not to give numerical 
predictions definitely. 

We  again would like to emphasize that an observation of 
$\mu^- N \rightarrow e^- N$ without observation of 
$\mu \rightarrow e +\gamma$ will be promising as 
a test of the present scenario. 

We hope that many physicists turn their attention 
to a possibility of the family gauge bosons with an
inverted mass hierarchy.

 \vspace{3mm}

{\Large\bf Acknowledgments} 

The author thanks T.~Yamashita  
for valuable and helpful conversations, 
especially, for pointing out a mistake in the earlier 
version of this work.
He also thanks H.~Yokoya for helpful comments on 
the direct production of $A_3^{\ 3}$, $A_3^{\ 2}$ and $A_3^{\ 1}$ 
at the LHC,  Y.~Kuno and H.~Sakamoto for useful comments on  
experimental status of $\mu$-$e$ conversion, M.~Koike for helpful 
comments on an estimate of the $\mu$-$e$ conversion, and 
M.~Tanimoto for valuable information on the recent estimates of 
$K^0$-$\bar{K}^0$ mixing. 


 \vspace{3mm}

{\Large\bf Appendix} 

In the K-Y model \cite{K-Y_PLB12}, charged lepton mass term is 
generated by a VEV of scalar $\Phi$ as follows:
$$
{\cal H}_{\rm Yukawa} = \frac{y_e}{\Lambda^2} \bar{\ell}_{Li} \Phi_i^{\ \alpha} 
\bar{\Phi}_\alpha^{\ j} e_{Rj} H_d  ,
\eqno(A.1)
$$
where $\ell_{Li} =(\nu_i, e^-_i)_L$.
Here, for simplicity, we have shown only the charged lepton term.  
For quark mass terms, we will need further scalars. 

On the other hand, family gauge boson masses are generated by 
another scalar $\Psi$:
$$
{\cal H}_{mass} = \frac{1}{2} \left( g_A  A_i^{\ k} \langle \Psi_k^{\ \alpha} \rangle
- g_B \langle (\Psi_i^{\ \beta}) \rangle B_\beta^{\ \alpha} \right) \left(
g_A \langle (\Psi^\dagger)_\alpha^{\ k} \rangle A_k^{\ i} -
g_B B_\alpha^{\ \gamma} (\langle \Psi^\dagger)_\gamma^{\ i} \rangle \right) + \cdots ,
\eqno(A.2)
$$
where $A$ and $B$ are gauge bosons of U(3) and U(3)$'$, respectively. 
The term ``$+ \cdots$" denotes other contributions except for that from 
$\Psi = ({\bf 3}, {\bf 3^*})$ of  U(3)$\times$U(3)$'$, i.e. 
contribution from $\Phi=( {\bf 3}, {\bf 3^*})$  
which gives quark and lepton masses,  
contribution from $\Psi' = ({\bf 1}, {\bf 6})$ which plays a role 
in giving $|M_{ij}(B)|^2 \gg | M_{ij}(A) |^2$,
and so on. 
When we assume a VEV form  
$$
\langle \Psi_i^{\ \alpha} \rangle = \delta_i^{\ \alpha} v_i ,
\eqno(A.3)
$$
we obtain
$$
{\cal H}_{mass} =\frac{1}{4} \sum_{i,j} (|v_i|^2 +|v_j|^2) 
\left( g_A A_i^{\ j} -g_B B_i^{\ j} \right)^2
+ \cdots . 
\eqno(A.4)
$$
Since we suppose $|\langle \Phi\rangle |^2 \ll |\langle \Psi\rangle |^2 \ll
|\langle \Psi' \rangle |^2$, i.e. in the limit of large masses of the gauge 
bosons $B_\alpha^{\ \beta}$,  we obtain
$$
M^2_{ij} \simeq \frac{g_A^2}{2} (|v_i|^2 + |v_j|^2 ) .
\eqno(A.5)
$$
(Hereafter, we denote $g_A$ as $g_F$.) 

In Eq.(A.5), since contributions of scalars which generate quark and lepton 
masses are very small, the contributions have been neglected. 
Hereafter, we neglect small contributions denoted by ``$+\cdots$". 
Under this approximation, family gauge bosons interact with quarks 
and leptons as follows:
 $$
{\cal H}_{fam} \simeq \frac{g_{F}}{\sqrt{2}} \left[ (\bar{e}_i \gamma_\mu e_j) 
+ (\bar{\nu}_i \gamma_\mu \nu_j) 
+ U^{* u}_{ik} U_{jl}^u (\bar{u}_k \gamma_\mu u_l )
+  U^{* d}_{ik} U_{jl}^d(\bar{d}_k \gamma_\mu d_l)  \right] (A_i^{\ j})^\mu .
\eqno(A.6)
$$
On the flavor basis in which the charged lepton mass matrix $M_e$ is diagonal, 
the family gauge bosons $A_i^{\ j}$ are in the eigenstates of mass. 
On the other hand, quark mass matrices $M_u$ and $M_d$ are, in general, not 
diagonal. 
$U^u$ and $U^d$ in Eq.(A.6) are mixing matrices which are described in the diagonal
basis of $M_e$. 
Therefore, there is no family number violation in the charged lepton sector, 
while family number violations can appear in the quark sectors through the mixings
$U^u$ and $U^d$. 

The interaction form (A.6) is always correct in a model in which mass matrices
of the charged leptons and family gauge bosons can be diagonalized simultaneously.

%

%

\end{document}